# Estimating ECG Intervals from Lead-I Alone: External Validation of Supervised Models

Ridwan Alam, *Member, IEEE*, and Collin M. Stultz, *Member, IEEE*

**Abstract— The diagnosis, prognosis, and treatment of a number of cardiovascular disorders rely on ECG interval measurements, including the PR, QRS, and QT intervals. These quantities are measured from the 12-lead ECG, either manually or using automated algorithms, which are readily available in clinical settings. A number of wearable devices, however, can acquire the lead-I ECG in an outpatient setting, thereby raising the potential for out-of-hospital monitoring for disorders that involve clinically significant changes in ECG intervals. In this work, we therefore developed a series of deep learning models for estimating the PR, QRS, and QT intervals using lead-I ECG. From a corpus of 4.2 million ECGs from patients at the Massachusetts General Hospital, we train and validate each of the models. At internal holdout validation, we achieve mean absolute errors (MAE) of 6.3 ms for QRS durations and 11.9 ms for QT intervals, and a MAE of 9.2 ms for estimating PR intervals. Moreover, as a well-defined P-wave does not always exist in ECG tracings – for example, when there is atrial fibrillation – we trained a model that can identify when there is a P-wave, and consequently, a measurable PR interval. We validate our models on three large external healthcare datasets without any finetuning or retraining - 3.2 million ECG from the Brigham and Women's Hospital, 668 thousand from MIMIC-IV, and 20 thousand from PTB-XL - and achieve similar performance. Also, our models significantly outperform two publicly available baseline algorithms. This work demonstrates that ECG intervals can be tracked from only lead-I ECG using deep learning, and highlights the potential for out-of-hospital applications.**

## I. INTRODUCTION

INTERVALS between two fiducial points on an electrocardiogram (ECG) provide crucial information about the electrical activity of the myocardium. Action potentials, spontaneously initiated at the sinoatrial node of the heart, traverse the heart via a specialized conduction pathway. A 12-lead ECG captures the cumulative electrical dipole across the heart as a continuous waveform consisting of P, Q, R, S, and T waves [1]. The temporal distances among the peak, onset, and offset of these waves - i.e., the fiducial points - provide information about the propagation of action potentials within different portions of the conduction system. Indeed, these temporal distances are often diagnostic of different cardiovascular disorders [2-6]. For example, during a $1^{st}$ degree atrioventricular (AV) block, the electrical conduction between the atria and the ventricles is slowed down at the AV node, and the PR interval - the distance between the P-wave onset and QRS-complex - increases over 200 ms. Such blocks can deteriorate to $2^{nd}$ degree or even complete heart blocks, leading to syncope, cardiac arrest, and stroke [3]. On the other spectrum, shortening of the PR interval below 120 ms predicts junctional rhythms and pre-excitation (see Fig. 1). Similarly, narrow QRS durations are associated with cardiac disorders such as atrial flutter and junctional tachycardia, whereas broad QRS complexes are evident of ventricular pacemakers and associated with hyperkalemia and bundle branch blocks [4]. QT prolongation is used as distinctive biomarkers for hypokalemia, hypocalcemia, myocardial ischemia, even for antiarrhythmic drug effect monitoring (see Fig. 1), and is predictive of a life-threatening arrhythmia Torsade-de-Pointes [5,6]. Thus, ECG intervals, solo or in combination, possess significant clinical utility in diagnosing critical cardiac conditions, devising treatment plans, and preventing severe outcomes.

The standard 12-lead ECG captures the time-dependent propagation of cardiac action potentials along twelve axes: leads I, II, III, aVL, aVR, aVF, V1-V6. For example, lead-I captures the conduction activity along the horizontal axis from right to left arm. Characteristic changes in the ECG intervals are not equally prominent across all the leads and vary with cardiac conditions [1]. For example, normal P-waves are best visible in the inferior leads - II, III, aVF - and are therefore considered best for measuring the PR interval. Clinical ECG machines also use multiple leads to generate reliable estimates of the ECG intervals, and it is still an open question whether reliable estimates of intervals can be obtained with lead-I ECG alone [7,8]. As many wearable outpatient ECG monitors can acquire lead-I and not all 12 leads, estimating intervals from lead-I ECG can enable numerous outpatient applications.

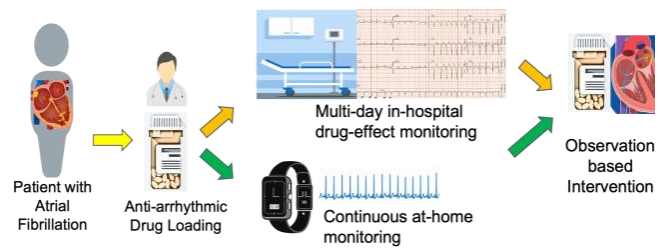

Figure 1. Example use-case of ECG interval monitoring. Clinical regimens to start some anti-arrhythmic drugs require the patient to stay in the hospital for 2-3 days, closely monitor their QT intervals for signs of potential adverse drug-effects, and intervene accordingly. QT monitoring from lead-I ECG only can enable similar care at home settings reducing hospitalization needs.

Recent rise in adoption of wearable and pocket ECG devices raise the possibility of outpatient cardiac monitoring by tracking ECG intervals in carefully selected patients [9,10]. Such applications require a reliable method for estimating intervals from ambulatory ECG recordings. Inspired by the success of deep learning approaches for predicting a variety of clinically meaningful outcomes from the 12-lead ECG [11-21], several investigators have developed methods for automated interval estimation using clinical 12-lead and wearable ECG devices [22-24]. Deep learning (DL) methods have shown robust performance against data variation and noise artifacts. Yet, generalization of such methods beyond the data used to train and test these models remains a critical hurdle in demonstrating the utility of such solutions in general populace outpatient settings [25,26]. In particular, algorithms built for single-lead ECG often struggle to generalize to data from different healthcare settings [27].

We aim to estimate the ECG intervals from lead-I ECG using deep learning. We propose three regression pipelines for estimating the PR, QRS, and QT intervals. Using a Resnet-based convolutional backbone, which we named IKres (lead-I EKG Resnet), we train three regression models, IKres-PR, IKres-QRS, and IKres-QT using an internal ECG dataset from the Massachusetts General Hospital (MGH). As a number of ECGs do not have a readily identifiable P-wave, as in the case of atrial fibrillation and atrial systole, we also develop a classification model, IKres-PRchk, that can identify when the P-wave exists in an ECG and consequently when there is a measurable positive PR interval. The resulting models are performant relative to existing models and algorithms on the internal-holdout dataset. Moreover, we validate the generalizability of these models on three large external datasets; from the Brigham and Women's Hospital, MIMIC-IV [28], and PTB-XL [29]. On these data, all proposed models demonstrate robust performances, similar to those achieved on internal data. These results highlight the potential of lead-I ECG in monitoring patients with cardiac risks in outpatient settings.

## II. RELATED WORKS

DL models for ECG analysis have notably advanced in recent years in identifying many cardiac arrhythmias and abnormalities, as well as in capturing the relationship between the ECG and hemodynamic parameters. Diagnosis and prognosis problems are often posed as classification tasks. For example, convolutional neural networks-based supervised classifiers, including residual neural networks (Resnet), have been proposed to predict various arrhythmias [11], atrial fibrillation [12], cardiac arrest [13], heart defects [14,15], heart blocks, and conduction disorders [16,17]. Scarcity of clinically validated annotations, especially for rare cardiac conditions and diseases, poses a perpetual challenge for supervised learning-based methods. Moreover, data quality, reliability, interrater variability, and confounders associated with health conditions add to the list of challenges. Beyond the supervised classifiers, self-supervised learning has been investigated to bypass the need for large datasets with clinically validated annotations. For ECG, representations learned this way have proven useful for downstream classification tasks, e.g., arrhythmia classification [18], atrial fibrillation and ventricular hypertrophy prediction [19]. Such representations are applied for building decision support systems in hypertension diagnosis [20]. Adapting contrastive learning methods, [21] explores the utility of such methods in identifying different cardiac disorders on the PTB-XL dataset. Being motivated by the ability to learn from 12-lead ECG, similar models are built and evaluated on single or four-leads ECG inputs from wearable or ambulatory devices [27,30,31]. Irrespective of the applications or the learning paradigms, these studies lend credence to the feasibility of Resnet as an attractive architecture for modeling ECG-based learning tasks. Hence, we build our models on IKres, a 1-d Resnet backbone, and adapt for regression of continuous variables.

While classification models have shown promise, research on regression tasks remains mostly uncharted [32]; especially whether DL can accurately estimate continuous variables such as the ECG intervals is an open question [33]. Few works have used all 12 leads to estimate the intervals. For example, [22] trained a 12-lead ECG-based Resnet model to estimate PR intervals on a selected subset of patients; excluding those with atrial fibrillation and other cardiac conditions. Some studies even posed the regression task differently as a classification mold. In [23], a two-lead (I and II) ECG model was trained to measure the QT interval as a quantized variable using multi-class classification. This model was then applied on a pocket ECG monitoring system, mECG, to predict cardiologist over-read corrected QT values (QTc) for 686 patients with genetic heart disease, where half of these patients had long QT syndrome. This approach achieved a standard deviation of the error (SDerr) of 23 ms for estimating the QTc intervals from the clinical ECG and 25 ms on the pocket ECG monitor. In another approach, DL models were trained to delineate an ECG beat and the predicted fiducial points were used to calculate the intervals [24,33]. Using a Cardiologs-proprietary DL model built on the U-net architecture and trained on Holter ECG, [24] tried to delineate ECG acquired from smartwatch and calculate the intervals. The best performance reported for this approach was SDerr of 22 ms and correlation coefficient of 0.57 in estimating QTc. No DL solution has been proposed yet for estimating QRS duration from ECG. The regression results of these early DL methods show scopes for improvement, yet their attempts to use DL models trained on clinical ECG to outpatient wearable or smartwatch ECG provide strong guidance on generalizability.

For deep-learning in healthcare, generalization remains the most-wanted yet the least-explored feature toward real-world impact. Most performant solutions reported in the literature use conspicuously small test-sets. For example, [15] tested their model only on 328 ECG from 328 patients, while their model was trained on 91 thousand samples from 54 thousand patients;

<div align="center">

TABLE I

DEMOGRAPHICS AND DATA STATISTICS

</div>

| Dataset | MGH | BWH | MIMIC-IV | PTB-XL |
|---|---|---|---|---|
| **Patients** | 903,593 | 667,157 | 155,481 | 17,379 |
| **ECG** | 4,223,689 | 3,170,600 | 668,697 | 19,705 |
| **Age (yr)** | 61 ± 18.7 | 60 ± 16.4 | 62.3 ± 17.8 | 61 ± 29.5 |
| **Female (%)** | 43.0 | 50.2 | 50.3 | 48.2 |
| **HR (bpm)** | 77 ± 20.3 | 77 ± 18.7 | 76 ± 18 | 73 ± 14.8 |
| **QT (ms)** | 394 ± 49.9 | 396 ± 47.6 | 398 ± 46.3 | 398 ± 36.2 |
| **QRS (ms)** | 97 ± 24.6 | 95 ± 22.2 | 97 ± 20.4 | 96 ± 17.7 |
| **PR (ms)** | 158 ± 43.9 | 160 ± 41.1 | 163 ± 33.5 | 166 ± 29 |

[18] had a test-set of 827 samples in contrast to 2.3 million training samples. Self-supervised pretraining is expected to learn and generate useful representations even for external datasets, though such generalizability remains under active exploration. [21] evaluated their pretrained models on about 2000 ECG split across 71 labels (average 30 samples per class), and kept the evaluation set as part of the pretraining dataset. For regression, the Cardiologs-model evaluation in [24] used only 85 patients with Covid-19 to explore the utility of the solution.

DL solutions for ECG interval regression from only lead-I ECG are at early development stages [34,35]; generalizability of such models to multi-hospital settings is critical for real impact. We propose this work as a baseline for related future research, especially on publicly available ECG datasets.

## III. DATA DESCRIPTION

We develop our method using data from one healthcare institution and evaluate the generalizability of these models on data from three independent "external" healthcare institutions, as described in Table I and Fig 2. Though these datasets contain 12-lead clinical ECG, we use only the lead-I ECG for this study.

### A. MGH-dataset

The *MGH-dataset* contains 4,223,689 clinical ECG records from 903,593 patients at the Massachusetts General Hospital (MGH) at Boston, MA, acquired between 1981 and 2020. Heart rate, PR, QRS, and QT intervals for each ECG are stored with other metadata in the dataset. These features were measured by the ECG acquisition machines (GE and Philips) and reviewed by attending cardiologists. The ECG recordings were acquired in millivolts (mV) of voltages with 12-bit quantization and at sampling rates of 250 Hz or 500 Hz. The heart-rate was labeled in beats-per-minute (bpm), the PR, QRS, and QT intervals were measured in milliseconds (ms).

### B. BWH-dataset

The *BWH-dataset* contains 3,170,600 ECG recordings from 667,157 patients who received care at the Brigham & Women's Hospital (BWH) at Boston, MA. The ECG and corresponding labels have similar acquisition characteristics (sampling rates, voltage levels, and interval units) as those from MGH, similar clinical equipment was used in collecting those data.

### C. MIMIC-IV-ECG

As part of the larger MIMIC-IV clinical database, the *MIMIC-IV-ECG* module [28] is hosted on Physionet [36] and contains 668,697 clinical ECG acquired between 2008 and 2019 from 155,481 patients at the Beth Israel Deaconess Medical Center (Boston, MA). The ECG acquisition machines were mostly from Burdick/Spacelabs, Philips, and GE. The machine-read ECG features, including intervals and fiducial points, and the summary reports are available for each ECG. Sampling rate for the ECG waveforms are 500 Hz, with other characteristics similar to those for the MGH dataset.

### D. PTB-XL

*PTB-XL* [29] is a large publicly available ECG dataset that contains 19,705 ECG from 17,379 patients with a variety of cardiovascular diagnoses including conduction disorders, myocardial infarctions, ischemia, and hypertrophic cardiomyopathy, as well as those without any cardiac disorders. The ECG data were acquired using devices by Schiller AG and are available at a sampling rate of 500 Hz. PTB-XL+[37] is an extension of this dataset annotating the intervals and the fiducial points using three algorithms, including the GE Marquette 12SL ECG analysis method [7].

The demographic properties of the four datasets are presented in Table I. The 'age' reported corresponds to that of a patient when the ECG was recorded; two ECGs from a patient on two different years have two different ages. Distributions of the PR, QRS, and QT intervals are shown in Fig 2.

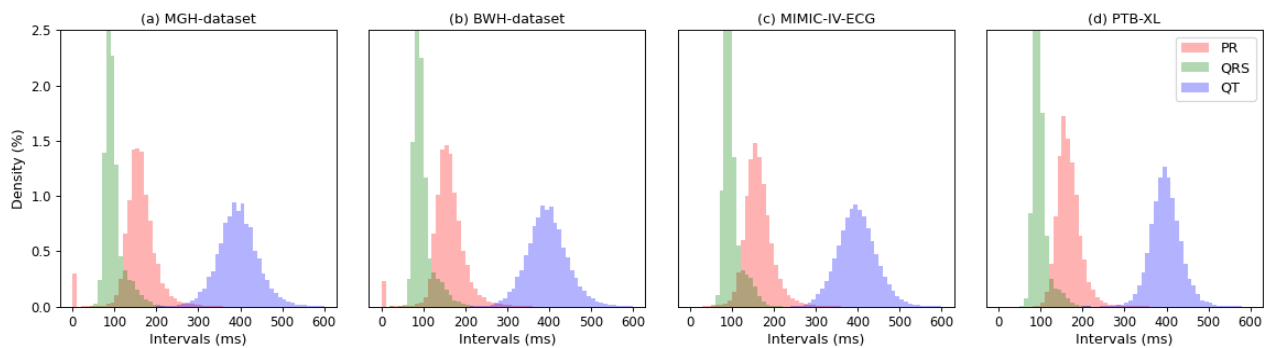

Figure 2. ECG intervals across hospitals show similar distributions. The variance is lower for QRS durations than PR or QT intervals. The P waves can be absent on the ECG of patients with certain conditions, causing non-normal spike at the zero-value for PR intervals.

## IV. METHODS

### A. Lead-I ECG and Intervals

Our objective is to estimate the PR, QRS, and QT intervals from a 10-second lead-I ECG recording. The sampling rates for the "input" lead-I ECG vary across the datasets, hence, for our proposed models, we resample the signal at 250 Hz to ensure uniformity. Then, we remove baseline wander and high frequency noise using a bandpass filter that only allows frequency components within a 0.05-to-40 Hz band. Following that, we exclude any 10-second ECG signal that has an absolute voltage amplitude larger than 5 mV. We intentionally avoid any signal normalization on the lead-I signal to preserve its comparative amplitude information. We use similar preprocessing methods to prepare the input signals for the baseline algorithms we use for comparison.

The ECG intervals (i.e., the labels) are measured by the proprietary algorithms built into the ECG machines. These algorithms often rely on multiple leads to first extract a median from the 10-second signals using peak detection and pattern matching. The median beat is delineated using heuristic rules implemented in the algorithm, and the intervals are calculated accordingly [7,8]. These intervals show similar distributions across the datasets, as shown in Fig 2. QRS and QT intervals over the large populations exhibit Gaussian-like distributions. But, for the PR interval, an additional mode is observed at zero for the two largest datasets, MGH and BWH. This phenomenon is due to instances when the P-wave cannot be clearly identified, in which cases the PR interval has a label of zero, similar observation has been reported in [22].

Given these distributions of the ECG intervals, we propose three separate regression models for the QT, QRS, and PR intervals, respectively. To address the challenges with the non-normality of the PR labels, we build a classifier to identify which ECG contains non-zero clinically sensible PR intervals. The labels for that classifier are binary; 1 representing instances when the PR interval is greater than zero.

### B. Proposed Models

We develop three deep learning regression models for three ECG intervals, PR, QRS, and QT, and a classification model to identify non-zero PR intervals, using the same backbone architecture, IKres (Fig. 3). IKres is a single-channel residual neural network (ResNet-18) consisting of four residual blocks. Each residual block is comprised of two convolutional layers and a skip connection. IKres takes in lead-I ECG and produces a representation tensor that can be fed into different projection heads for different tasks. To ingest the 10-second lead-I ECG signal as a single-channel input, we add a 1-d convolution layer followed by batch normalization and non-linear activation unit before the residual blocks. The input channel is of 1x2500 samples length, as we resample all 10-second ECG lead-I signals to 250 Hz sampling rate to get the input tensor. The ingest layer convolves this tensor with single sample stride over 64 filters. We used a kernel size of 16 units for all convolutional filters. For the four residual blocks, we learn convolution layers with 128, 196, 256, and 320 filters, consecutively. For each block, the skip connections are implemented with max pooling and 1-to-1 convolution layer. Batch normalization and rectified linear units follow each convolution layer of the model. Using average pooling, we get a 1x320 representation from the last residual block of IKres.

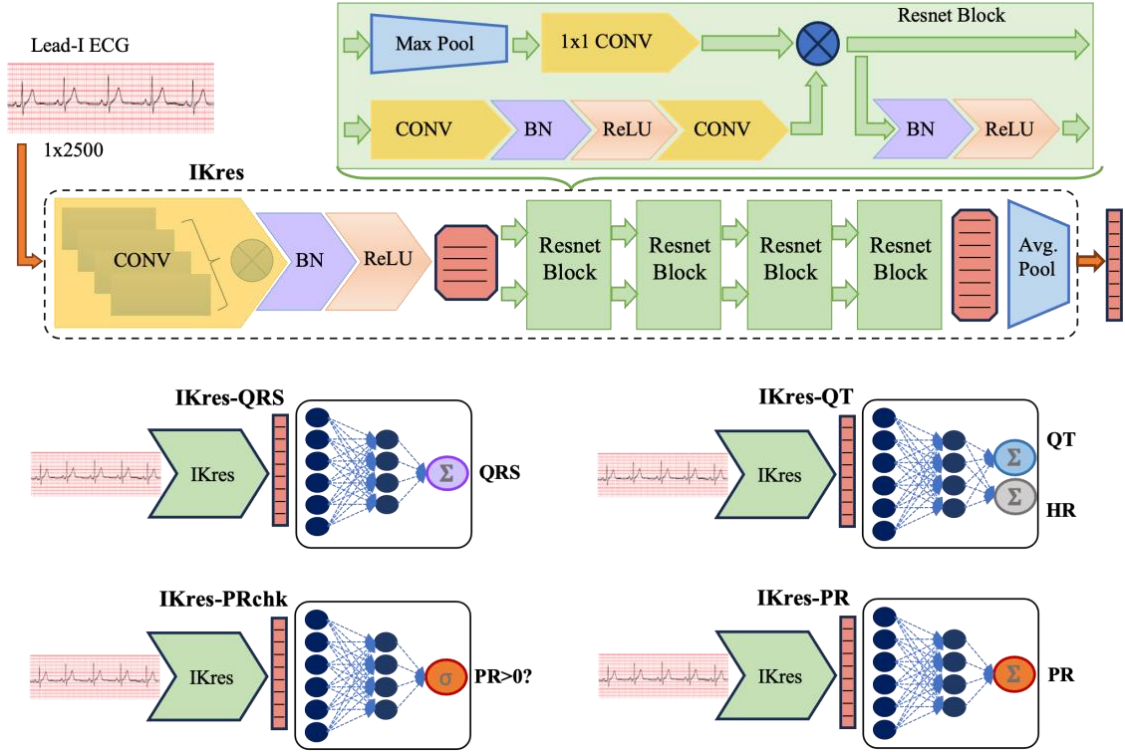

Figure 3. Model architectures. IKres is a modified Resnet-18 backbone that takes in single-channel 10-second lead-I ECG (sampled at 250 Hz) and outputs 1x320 embeddings. We train four separate but similar multilayer perceptron heads along with this IKres backbone to build four models for estimating the ECG intervals from the input lead-I ECG; IKres-QT estimates QT intervals and heart rates, IKres-QRS regresses QRS durations, IKres-PR estimates the PR intervals, and IKres-PRchk identifies non-zero PR intervals.

To build the three regression models, IKres-PR, IKres-QRS, and IKres-QT, and the classification model IKres-PRchk, we use similar multilayer perceptron-based regression and classification heads with the IKres backbone (Fig. 3). We pass the representation from the residual blocks on to two fully-connected dense layers. For IKres-PR and IKres-QRS, the corresponding dense layers output a single random variable as the estimates for PR interval and QRS duration, respectively. The dense layer for IKres-QT outputs two random variables as the estimate of QT interval and heart rate. And, for classification task, the IKres-PRchk has a sigmoid layer after the dense layers, which outputs logit values as the probability of the PR interval having a clinically sensible non-zero value.

We use PyTorch framework to implement the architectures as well as for model training and evaluation.

### C. Model Training

The *MGH-training* set contains 70% of the MGH-dataset (3.06 million ECGs, arising from 653 thousand patients). A training-time *MGH-validation* set is used to determine when training should end and contains 15% of MGH-dataset (534 thousand ECG from 115 thousand patients), and the remaining 15% of the data comprises our *MGH-holdout* set (633 thousand ECGs from 136 thousand patients). The *MGH-holdout* set is used only for testing the model performances and is not seen by any model during training phases. Given that each patient generally has several ECGs, we ensure that all ECGs from a given patient only appeared in one of these subsets, i.e., the data splits have no overlap with respect to patient data. Again, although ECGs in each of these datasets contain all 12 leads, only data from lead-I is used to train and evaluate the models.

For training all four models, IKres-QRS, IKres-QT, IKres-PR, and IKres-PRchk, the initial model weights of the end-to-end pipeline, including the IKres backbone, were set according to the Kaiming initialization method with random variables from a normal distribution with variance depending on the layer size [40]. Though the same architecture backbone IKres is used for all four models, the training involves learning weights for the end-to-end pipeline, no pretrained components are used. For training, we minimize an objective cost function; for the regression models, the mean square error (MSE) between the predicted intervals and their corresponding labels is used as the cost function. The ECG intervals and the heart rates are normalized to zero-mean and unit-variance distributions (i.e., z-scored) during training, and the distribution means and variances from the *MGH-training* set is stored as model parameters. During inference, the model predictions undergo the inverse z-transformation using those means and variances to acquire the absolute values in their corresponding units. For the classification task, we use the binary cross-entropy (BCE) loss. The prevalence of zero-valued PR intervals is significantly low; hence, we employ weighted loss for training the classifier. Back-propagation with an ADAM optimizer is used to minimize the cost functions. For controlling the learning rate, we use a step scheduler to decay the rate in half every 3 epochs, starting from 0.01, and a batch size of 512 is used. Early-stopping is used to reduce the risk of overfitting based on the validation loss. Essentially, training ends when the MSE or the BCE starts to rise in the *MGH-validation* set.

### D. Baseline Algorithms

To compare the performance of our proposed methods for the ECG interval regression, we use two baseline ECG analysis algorithms for delineating the intervals and heart rate.

1) **NeuroKit**
   The NeuroKit library [38] provides algorithms for ECG analysis and is available online (http://github.com/neuro-psychology/NeuroKit). We use the signal quality measurement, peak detection, and ECG delineation algorithms from this library in building the interval calculation pipeline using lead-I ECG signals. The input signal is preprocessed with a bandpass filter that only keeps the 0.05-to-40 Hz frequency components, before performing the delineation. The built-in delineation algorithm first detects the R peaks and use heuristic parameters to segment specific location on the signal in detecting the other peaks. The intervals are calculated for each beat, corresponding to each detected R-peaks, and their averages are used as the estimates. This library is implemented in Python.

2) **ECGdeli**
   ECGdeli is also an open-source ECG delineation toolkit implemented in Matlab [39], and publicly available online (http://github.com/KIT-IBT/ECGdeli). We use the built-in Filtering and ECG_Processing modules of this library to build the pipeline for calculating PR, QRS, and QT intervals. The module builds on the fiducial points obtained by peak detection and heuristic windowing of the signal, and the features are computed separately for each available beat. The default feature extraction algorithms are designed to calculate the global mean of the intervals over all 12 leads of the ECG, if available. We update the pipelines such that the methods use only the lead-I ECG, keeping the inputs same as those for the proposed models. While the algorithm claim to be performant for any sampling rates, we observed significant deterioration in performance when the input sampling rate was 250 Hz. Hence, we had to resample the lead-I ECG signals from all four datasets to 500 Hz, as the ECGdeli algorithm apparently is robust only at that sampling rate.

### E. Evaluation

We evaluate the trained models on four datasets, the *MGH-Holdout*, the *BWH-dataset*, the *MIMIC-IV-ECG*, and the *PTB-XL* datasets, from four different healthcare institutes, as described in Section III. To quantify the residuals between the estimated intervals and their corresponding labels, we calculate mean absolute error (MAE) and the standard deviation of the error (SDerr). MAE quantifies the absolute difference between the estimation and its corresponding label and summarize over

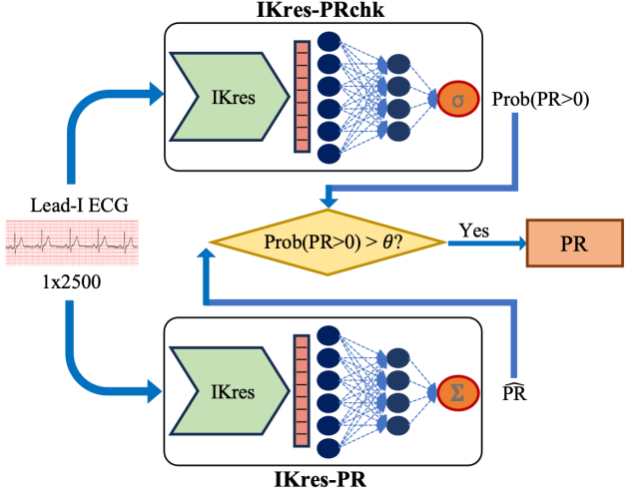

**IKres-PRchk**

Prob(PR>0)

Lead-I ECG

1x2500

Prob(PR>0) > θ? → Yes → PR

**IKres-PR**

PR

Figure 4. PR interval estimation from 10-sec lead-I ECG. During inference time, IKres-PRchk predicts the probability of the presence of a strictly positive PR interval in the input ECG. This probability is used to filter-out unreliable estimations by IKres-PR.

all estimations as the average of those differences. SDerr refers to the spread of the distribution of the differences on a Bland-Altman plot as presented on some related research [23,24]. We also report the Pearson's correlation coefficient (Pearson-R) for assessing the similarity between the estimated values and the labels. Using these metrics, we compare the proposed deep-learning solutions with the baseline algorithms.

For the estimation of PR intervals, we consider the possibility of any non-normality in the distribution (Fig. 2), which occurs when the P-waves are not present on the ECG. Toward that goal, we use the predictions from the classifier (IKres-PRchk) as a selector for the estimations from the IKres-PR model, as shown in Fig 4. For evaluating the classification task, we report the area under the receiver-operating-curve (AUC) and the area under the precision-recall curve (AUPRC). We find the cutoff or threshold - between 0 and 1 - from the MGH-training set to binarize the model predicted logits for classification; we choose the threshold value that corresponds to maximum specificity and sensitivity on that set. Using that threshold, we report accuracy, sensitivity, and specificity of IKres-PRchk on the internal holdout and the external datasets. Then, we present the estimation results on this subset of predictions using the same regression metrics, MAE, SDerr, and Pearson-R.

We use kernel density estimation (KDE) plots to visualize the regression performance of the models. The KDE plots use Gaussian kernels to present the distribution of the model predictions as continuous density estimates. Plotting these density contours against the ground truth interval labels, the correlation between the predictions and the labels as well as the alignment of the model with an ideal predictor can be visually compared.

## V. RESULTS & DISCUSSIONS

The performance of the proposed models in estimating the ECG intervals are presented here. We show the generalizability of the proposed models on unseen data from the same hospital

(*MGH-holdout*) and from three completely external hospitals (*BWH-dataset*, *MIMIC-IV-ECG*, and *PTB-XL*) using the same evaluation metrics. In all cases, we compare the proposed models against the baseline algorithms explained in Section IV.

### A. QRS Durations

The IKres-QRS model is trained on the *MGH-training* set to regress QRS durations from lead-I ECG. The performance of the model in this regression task is presented on Table II. While the baseline algorithms, NeuroKit and ECGdeli, achieve at best a mean absolute error (MAE) of about 20 ms on the *MGH-holdout* set, the estimations by the IKres-QRS are notably better approximations of the true duration values with an MAE of only 6.35 ms. Similarly, the model achieves an SDerr of 9.3 ms in comparison to about 20 ms from the baseline methods, highlighting the fact that the model estimations are less dispersed and more consistent. Not only in reducing the residual gaps, IKres-QRS also performs better than the baseline methods in increasing the correlation coefficients between the estimations and the true-values, achieving a 0.91 Pearson-R and 0.82 coefficient-of-determination, $R^2$; as shown in Table II. The IKres-QRS does not suffer any loss in performance across all three external datasets, two of which are publicly available, collected at different locations with different ECG machines.

### B. QT Intervals and Heart Rates

The average QT interval for a normal heart rhythm is about 400 ms, which represents more than 40% of the duration of an average heartbeat. Even on normal hearts, physical exertion or excitement can increase the heart-rate, leading to a decrement in the QT interval. Hence, in clinical practice, the QT interval is often "corrected" or adjusted with respect to the heart-rate [35]. Considering this clinical utility, we built the IKres-QT model to infer both the QT interval and the concurrent heart-rate from the lead-I ECG. The utility of such regression

### TABLE II
### QRS DURATION ESTIMATION

| METRICS | METHODS | DATASETS | | | |
|---|---|---|---|---|---|
| | | **MGH-Holdout** | **BWH** | **MIMIC-IV** | **PTB-XL** |
| | | N=564,613 | 3,170,600 | 668,697 | 19,705 |
| *MAE (ms)* | NeuroKit | 20.1 | 19.6 | 18.2 | 26.4 |
| | ECGdeli | 32.7 | 34.2 | 31.3 | 30.7 |
| | IKres-QRS | 6.3 | 6.6 | 6.4 | 6.4 |
| *SDerr (ms)* | NeuroKit | 31.0 | 30.2 | 26.7 | 31.0 |
| | ECGdeli | 21.1 | 21.3 | 20.2 | 18.3 |
| | IKres-QRS | 9.3 | 9.6 | 9.1 | 8.4 |
| *Pearson-R* | NeuroKit | 0.235 | 0.2291 | 0.284 | 0.315 |
| | ECGdeli | 0.496 | 0.479 | 0.486 | 0.480 |
| | IKres-QRS | 0.909 | 0.903 | 0.90 | 0.884 |

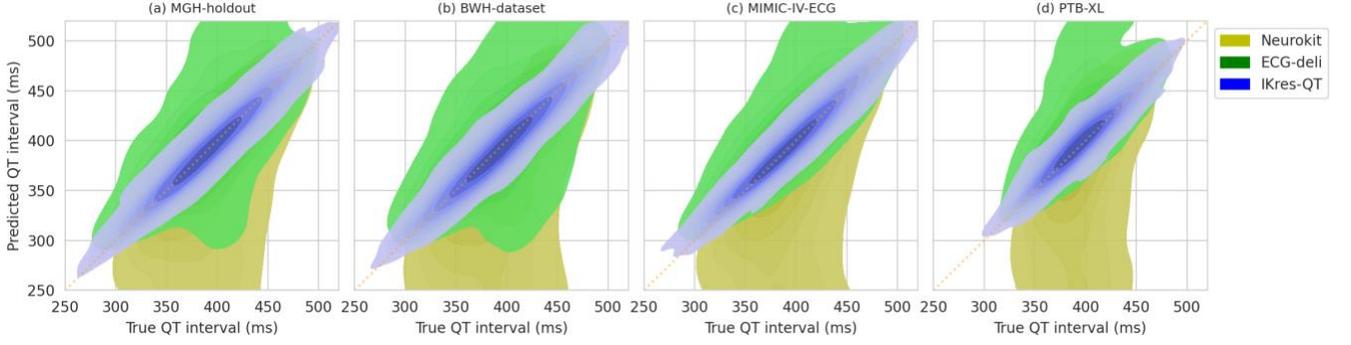

Figure 5. QT interval estimation with IKres-QT. Across four validation sets, the density plots of the inferred intervals against the true labels show that IKres-QT consistently outperforms the baseline algorithms. The orange dotted-lines represent ideal estimations.

models in clinical application can range from antiarrhythmic drug effect to monitoring the QT interval in patients with inherited QT-prolongation syndromes.

The heart-rate labels are ideally calculated as the number of QRS-complexes - i.e., the number of beats - present in an ECG signal over a minute long time window, and the QRS complexes are often the visibly predominant ECG waves. From Table III, IKres-QT can near-perfectly estimate the heart-rates with an MAE of only one beat-per-minute (bpm). Both the baselines, NeuroKit and ECGdeli, also similarly performs for this task. Notably, while ECGdeli performs relatively better on PTB-XL and MIMIC-IV-ECG, the same implementation of it drastically underperforms on the MGH and BWH data. This points to a potential defect of the published algorithm, suggesting further investigation on the underlying heuristics and assumptions.

In estimating the QT intervals, the IKres-QT achieves an MAE of 11.9 ms on the MGH-holdout set, shown in Table IV. Considering that the average QT interval is 400 ms, the mean absolute percentage error (MAPE) is less than 3%, which is a significant improvement over the methods reported in related literature. The difficulty of this task is evident from the poor performance by the baseline algorithms. The proposed model generalizes to the external datasets with similar performance; in fact, performs slightly better on the publicly available datasets. Moreover, the SDerr reported in [23,24] is about 25 ms, the proposed IKres-QT achieves an average 17.5 ms SDerr. The distribution of the estimated QT intervals for each validation dataset is presented in Fig 5 using KDE plots. The distribution of the QT intervals estimated by the IKres-QT model is comparatively better correlated to the interval labels, which is also evident from Table IV, as it achieves more than 90% Pearson correlation coefficient for all four datasets.

### C. PR Intervals

As described in Fig 4, the PR interval estimation involves a classifier and a regressor in tandem. The classifier IKres-PRchk identifies whether the input ECG has a positive PR interval. If a positive PR interval is predicted, then the regression model IKres-PR infers the PR interval for that input. Here, we first present the results of IKres-PRchk for the classification task –

#### TABLE III
##### HEART-RATE ESTIMATION

| METRICS | METHODS | DATASETS | | | |
|---|---|---|---|---|---|
| | | **MGH-Holdout** | **BWH** | **MIMIC-IV** | **PTB-XL** |
| | | N=564,613 | 3,170,600 | 668,697 | 19,705 |
| *MAE (bpm)* | **NeuroKit** | 1.78 | 1.77 | 1.36 | 1.15 |
| | **ECGdeli** | 9.25 | 14.32 | 0.93 | 0.87 |
| | **IKres-QT** | 1.15 | 1.12 | 1.03 | 1.08 |
| *SDerr (ms)* | **NeuroKit** | 4.99 | 4.86 | 4.21 | 3.55 |
| | **ECGdeli** | 570.9 | 729.0 | 3.19 | 2.59 |
| | **IKres-QT** | 2.75 | 2.33 | 2.12 | 1.64 |
| *Pearson-R* | **NeuroKit** | 0.965 | 0.966 | 0.972 | 0.971 |
| | **ECGdeli** | 0.035 | 0.028 | 0.984 | 0.985 |
| | **IKres-QT** | 0.990 | 0.992 | 0.993 | 0.994 |

#### TABLE IV
##### QT INTERVAL ESTIMATION

| METRICS | METHODS | DATASETS | | | |
|---|---|---|---|---|---|
| | | **MGH-Holdout** | **BWH** | **MIMIC-IV** | **PTB-XL** |
| | | N=564,613 | 3,170,600 | 668,697 | 19,705 |
| *MAE (bpm)* | **NeuroKit** | 81.3 | 90.8 | 97.0 | 78.8 |
| | **ECGdeli** | 34.7 | 37.5 | 29.9 | 30.8 |
| | **IKres-QT** | 11.9 | 12.3 | 10.8 | 10.7 |
| *SDerr (ms)* | **NeuroKit** | 82.1 | 85.6 | 86.5 | 78.8 |
| | **ECGdeli** | 51.0 | 54.3 | 42.4 | 42.9 |
| | **IKres-QT** | 18.9 | 18.6 | 16.7 | 15.8 |
| *Pearson-R* | **NeuroKit** | 0.390 | 0.368 | 0.343 | 0.320 |
| | **ECGdeli** | 0.575 | 0.534 | 0.689 | 0.569 |
| | **IKres-QT** | 0.919 | 0.922 | 0.933 | 0.904 |

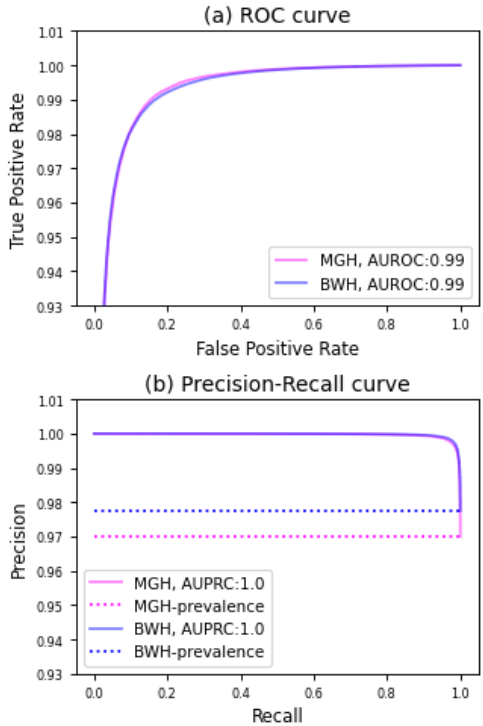

| METRICS | METHODS | DATASETS | | | |
|---|---|---|---|---|---|
| | | MGH-Holdout | BWH | MIMIC-IV | PTB-XL |
| | | N=528,446 | 2,973,732 | 643,287 | 18,831 |
| *MAE (bpm)* | NeuroKit | 35.4 | 35.9 | 36.9 | 49.0 |
| | ECGdeli | 66.4 | 71.2 | 65.9 | 72.8 |
| | IKres-PR | 9.2 | 9.1 | 12.0 | 8.6 |
| *SDerr (ms)* | NeuroKit | 33.7 | 33.2 | 31.8 | 35.3 |
| | ECGdeli | 63.1 | 83.5 | 28.7 | 24.1 |
| | IKres-QRS | 15.9 | 15.3 | 18.8 | 12.1 |
| *Pearson-R* | NeuroKit | 0.464 | 0.472 | 0.497 | 0.298 |
| | ECGdeli | 0.314 | 0.244 | 0.630 | 0.604 |
| | IKres-PR | 0.873 | 0.882 | 0.818 | 0.892 |

Figure 6. IKres-PRchk identifies positive PR intervals from lead-I ECG. 3% of the ECG from the *MGH-holdout* set and 2.3% of those from *BWH* has zero-labeled PR intervals (the negative class for the classifier), the other two datasets do not have this category.

detecting the presence whether a positive PR interval exists.

The performance of this classifier in identifying the positive PR intervals vs the non-positive intervals is presented in Fig 6. As presented in Fig 2, only the MGH and BWH datasets have significant presence (3% and 2.3%, respectively) of such outlier labels. IKres-PRchk achieves 0.99 AUC and 1.0 AUPRC on both datasets. As we find the classification threshold from the MGH-training set to binarize the predicted logits, we find the threshold value that corresponds to maximum specificity and sensitivity on that set. Using that threshold of 0.43, we achieve 99.5% specificity and 78% sensitivity on the MGH-holdout set, with an accuracy of 78.8%. For the BWH-dataset, we achieve 95.3% specificity and 95.8% sensitivity. As the negative class refers to the non-positive PR intervals, the high specificities on both datasets indicate the notable performance of IKres-PRchk in filtering out ECG with non-positive PR intervals.

The positive PR intervals are estimated by IKres-PR as a regression task on the subset of the validation sets, as identified by the classifier (Fig. 4). The PR estimation performance of this method is presented on Table V and Fig 7. Here, the estimations by the heuristic algorithms such as NeuroKit and ECGdeli suffer MAE of around 36 ms and 66 ms, respectively. Reliance on localized heuristics for identification of the fiducial points, namely P-onset and QRS-onset, may have led to such lower performance. On the other hand, IKres-PR learns from the millions of ECG examples with positive PR intervals to estimate the PR intervals from the input lead-I ECG. The MAE of the model predictions to the interval labels is only 9.2 ms on the MGH-holdout set. More importantly, the IKres-PR demonstrates similar generalizability as the other proposed interval estimation models. The estimations by IKres-PR on BWH-dataset, MIMIC-IV-ECG, and PTB-XL show an average MAE of 9.9 ms. For an average 'normal' PR interval of 160 ms, a 9 ms absolute error corresponds to a percentage error of 5.6%. Similarly, the SDerr remains steady across the four datasets with an average deviation of 15.5 ms. Notably, the correlation coefficients improve to 87% and 88%, respectively, for MGH-holdout and BWH-dataset, compared to the 60% correlation that we achieve in a setup without considering the positive PR identifications by IKres-PRchk. We present this estimation performance on Fig 7 using the KDE plots for both without and with the PR interval filtering that we implemented using the classifier. This figure visually explains the process of eliminating the examples where the PR interval labels are zero. For the MIMIC-IV-ECG and PTB-XL, this approach does not contribute much as such labels are not present in those datasets.

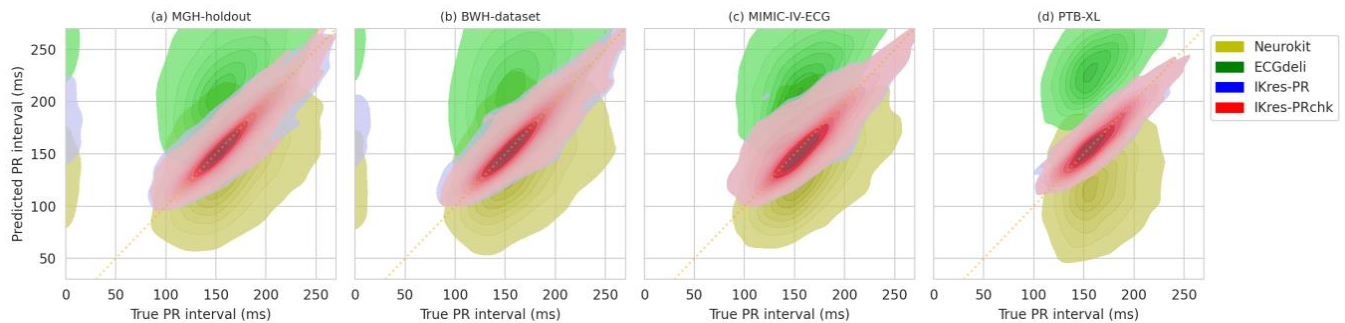

Figure 7. PR interval estimation from lead-I ECG. On the MGH and BWH datasets, the density plots show that IKres-PR estimates positive PR intervals even when the labels are zero. IKres-PRchk can identify such anomalies and thus improve joint performance.

## VI. Conclusion

Clinically, automated ECG interval estimation is critical for monitoring and assessing patient conditions in numerous cardiovascular disease diagnosis and treatment planning [2], but the current state of such care is only limited to hospital settings. The ability to track these ECG intervals from lead-I only can enable these use-cases to outpatient ambulatory settings [9,10]. We hypothesize that the lead-I ECG contains sufficient information to reliably estimate the ECG intervals that would have been obtained by analysis of the 12-lead ECG, and deep learning can extract and utilize that information. We present example DL models to estimate the PR, QRS, and QT intervals and report results that can be used as baselines for future research. For training and evaluating the proposed IKres-PR, IKres-QRS, and IKres-QT models, we use the intervals that was generated by clinical ECG machines and signed off by cardiologists as the labels or ground truths. These intervals are measured by the proprietary (GE/Phillips) algorithms using all 12 leads. The models learn to estimate those intervals from lead-I ECG only, even though lead-I is not the best electrical axis to capture the cardiac repolarization, but the best suitable option to generalize beyond inpatient care setting. This performance demonstrates strong support for that hypothesis and highlights the potential of estimating accurate ECG intervals from ambulatory wearable ECG devices.

A major challenge for such deep-learning solutions is their struggle to generalize beyond the internal data. We train our models only on the single-lead ECG signal, without any source-dependent modification - e.g., normalization - nor with any metadata inclusion. Thus, the models are "blind" to any hospital or patient-specific information. The resulting generalizability of the models is validated on data from three other hospitals. Beyond our internal holdout test data from MGH, we validate the models on external data from BWH, MIMIC-IV-ECG, and PTB-XL. The performance by the models in estimating the ECG intervals remains reliably consistent across these datasets, and better than those from recent studies [23,24] and those from baseline algorithms.

We report a phenomenon observed in the distribution of PR intervals, especially on large ECG datasets from hospitals with millions of patients. Such non-normality, if not properly addressed, impacts model performance and generalizability. In this work, we use a classifier-regressor tandem method to account for such anomaly. Moreover, the voltage amplitude of a P-wave is much lower on all ECG leads in comparison to the R and T waves. Certain cardiac conditions or drug interactions can decrease those more, making estimation of PR intervals, even for an expert electrophysiologists, very challenging. Reliance on localized heuristics by existing algorithms, such as NeuroKit and ECGdeli, for identification of the fiducial points, namely P-onset and QRS-onset, leads to poor performance. In comparison, careful data selection for regression model training on only positive clinically sensible PR intervals lead IKres-PR and the implementation of IKres-PRchk to identify non-positive intervals achieve reliable performance on four evaluation sets.

Among the limitations and future prospects of this work, the general applicability of these models across ethnicity, race, and age is yet to be explored. Also, neither of the four datasets were acquired with wearable ECG devices. Hence, our approach requires further prospective validation in ambulatory settings and with various wearable ECG monitors. While we explored many DL architectures for this task, the field is constantly improving with novel architectures showing advanced capabilities in many applications. Hence, we highlight the detail of the used architectures with the flexibility to update the backbone of the proposed models as better architectures are explored in the future.